\documentclass[aps,prd,groupedaddress]{revtex4}
\usepackage{graphicx,epsf,amssymb,amsmath,latexsym}

\newcommand{\ct}[1]{\cite{#1}}

\newcommand\pa{\partial}

\newcommand\be{\begin{equation}}
\newcommand\ee{\end{equation}}

\newcommand\lab[1]{\label{eq:#1}}
\newcommand\rf[1]{(\ref{eq:#1})}

\begin{document}

\title{Axions and Photons In Terms of ``Particles'' and ``Anti-Particles''}

\vspace{.3in}

\author{Eduardo I. Guendelman}
\email{guendel@bgu.ac.il}
\author{Idan Shilon}
\email{silon@bgu.ac.il}

\affiliation{Physics Department, Ben-Gurion University of the Negev, Beer-Sheva 84105, Israel}

\vskip.3in

\begin{abstract}

The axion photon system in an external magnetic field, when for example considered with the 
geometry of the experiments exploring axion photon mixing (which can be represented by a 
1+1 effective model) displays a continuous axion-photon duality symmetry in the limit the 
axion mass is neglected. The conservation law that follows from this symmetry is obtained. 
The magnetic field interaction is seen to be equivalent to first order to the interaction of a 
complex charged field with an external electric potential, where this ficticious "electric 
potential" is proportional to the external magnetic field. This allows one to solve for the 
scattering amplitudes using already known scalar QED results. Axion photon oscillations can 
be understood as violations of a charge symmetry in the scalar QED language. Going beyond 
the linear theory, the axion photon system in a self consistent magnetic field is shown, using 
this formalism, to have interesting  soliton solutions that represent new non gravitational ways 
of trapping light. Finally, generalizing the scalar QED formalism to 2+1 dimensions makes it 
clear that a photon and an axion splitt into two components in an inhomogeneous magnetic 
field, an effect that reminds us of the Stern Gerlach experiment. 
 
 \end{abstract}

\maketitle

\setcounter{equation}{0}

\section{Introduction}

The possible existence of a light pseudo scalar particle is a very interesting possibility. For example,
the axion \cite{Peccei}, \cite{weinberg}, \cite{Wilczek}  which was introduced in order to solve the strong CP problem has since then also been postulated as a candidate for the dark matter.
A great number of ideas and experiments for the search this particle have been proposed
\cite{Goldman}, \cite{Review}.

Here we are going to focus on a particular feature of the axion field $\phi$: its coupling to the photon 
through an interaction term of the form $g \phi \epsilon^{\mu \nu \alpha \beta}F_{\mu \nu} F_{\alpha \beta}$.
In fact, a coupling of this sort is natural for any pseudoscalar interacting with
electromagnetism, as is the case of the neutral pion coupling to photons (which, as a consequence of this interaction decays into two photons).

It was recognized by Sikivie that axion detection exploiting axion to photon
conversion in a magnetic field was a possibility \cite{Sikivie}.

A way to explore for observable consequences of the coupling of a light scalar 
to the photon in this way is to subject a beam of photons to a very strong magnetic field. This affects the optical properties of light which could lead to testable consequences \cite{PVLAS}. Also, a magnetic field in the early universe can lead to interesting photon-axion conversion effects \cite{Yanagida} and in the laboratory photon-axion conversion effects could be
responsible for the "light shining through a wall phenomena ", which are obtained by first
producing axions out of photons in a strong magnetic field region, then subjecting the mixed beam of photons
and axions to an absorbing wall for photons, but almost totally transparent to axions due to their weak 
interacting properties which can then
go through behind this "wall", applying then another magnetic field one can recover once again some photons 
from the produced axions \cite{LSTW}, \cite{Rabadan}. Notice however that the "light shining through a wall phenomena "
involves two interactions, once to produce the axions and then to obtain photons once again from the produced axions.
Since the axion photon coupling is so small, the amplitude for such effect is highly suppressed.

Here we wish to study properties of the axion-photon system in the presence of a strong magnetic field. By representing axions and photons as particles and anti particles, one can obtain known results and and we will show also that photons and axions splitt in the presence of an external magnetic field, in a way that we will make more precise. By this we mean that  from a beam of photons we will get two different kinds of scattered components (plus the photons that do not suffer any interactions), each of the scattered beams has also an axion component, but each of the beams is directly observable due to its photon component and an observable process is obtained to first order in the axion photon interaction (unlike the ``light shining through a wall''  phenomena). Although we cannot claim yet that this could provide a more favorable experimental set up, since such subject involves many practical questions in addition to the existence of the first order process. Beyond the question of observability, the existence of this kind of effects highlights many basic features of the axion photon system.

As another consequence of the interaction term, we will also show that when considering the effects which result from taking into account the back reaction of the axions and electromagnetic waves on the strong ``external'' magnetic field we can obtain localized soliton like solutions. We will take the direction of the self consistent mean field as orthogonal to the direction of propagation of the axion and of the electromagnetic component with rapid time dependence but parallel to the polarization of the electromagnetic waves. The resulting planar soliton solutions appear to be stable since they produce a different magnetic flux than a configuration where the localized axion photon configuration is absent, that is, when we have just the a constant magnetic field all over space (our choice of stable vacuum).

\section{Action and Equations of Motion}

The action principle describing the relevant light pseudoscalar coupling to the photon is

\begin{equation}
\label{axion photon ac }
S =  \int d^{4}x 
\left[ -\frac{1}{4}F^{\mu\nu}F_{\mu\nu} + \frac{1}{2}\partial_{\mu}\phi \partial^{\mu}\phi - 
\frac{1}{2}m^{2}\phi^{2} - 
\frac{g}{8} \phi \epsilon^{\mu \nu \alpha \beta}F_{\mu \nu} F_{\alpha \beta}\right].
\end{equation}

We now specialize to the case where we consider an electromagnetic field with propagation along the $y$ and $z$ directions and where a strong magnetic field pointing in the $x$-direction is present. This field may have an arbitrary space dependence in $y$ and $z$, but it is assumed to be time independent. In the case the magnetic field is constant, see for example \cite{Ansoldi} for general solutions.

For the small perturbations, we consider only small quadratic terms in the action for the axion and the electromagnetic fields, following the method of (for example) Ref. \cite{Ansoldi}, but now considering a static magnetic field pointing in the $x$ direction having an arbitrary $y$ and $z$ dependence and specializing to $y$ and $z$ dependent electromagnetic field perturbations and axion fields. This means that the interaction between the background field , the axion and photon fields reduces to
 
\begin{equation}
\label{axion photon int }
S_I =  - \int d^{4}x 
\left[ \beta \phi E_x \right],
\end{equation}

where $\beta = gB(y,z) $. Choosing the temporal gauge for the photon excitations and considering only the $x$-polarization for the electromagnetic waves (since only this polarization couples to the axion) we get the following 2+1 effective dimensional action (A being the x-polarization of the photon, so that $E_x = -\pa_{t}A$)

\begin{equation}
\label{2 action}
S_2 =  \int dydzdt 
\left[  \frac{1}{2}\partial_{\mu}A \pa^{\mu}A+ \frac{1}{2}\partial_{\mu}\phi \pa^{\mu}\phi - 
\frac{1}{2}m^{2}\phi^{2} + \beta \phi \pa_{t} A \right].
\end{equation}

Since we consider only $A=A(t,y, z)$, $\phi =\phi(t,y,z)$, we have avoided the integration over $x$. For
the same reason $\mu$ runs over $t$, $y$ and $z$  only . This leads to the equations

\begin{equation}
\label{eq. ax}
\pa_{\mu}\pa^{\mu}\phi + m^{2}\phi =  \beta \pa_{t} A
\end{equation}

and

\begin{equation}
\label{ eq. photon}
\pa_{\mu} \pa^{\mu}A = - \beta \pa_{t}\phi.
\end{equation}

As is well known, when choosing the temporal gauge the action principle cannot reproduce the Gauss 
constraint (here with a charge density obtained from the axion photon coupling) and has
to be imposed as a complementary condition. However this 
constraint is automatically satisfied here just because of the type of dynamical reduction
employed and does not need to be considered  anymore.

\section{The Continuous Axion Photon Duality Symmetry and the Scalar QED analogy}

Without assuming any particular $y$ and $z$-dependence for $\beta$, but still insisting that 
it will be static, we see that in the case $m=0$, we discover a continuous axion 
photon duality symmetry (these results were discussed previously in the 1+1 dimensional case, where only $z$ dependence was considered in \cite{duality}), since

\begin{enumerate}

  \item  The kinetic terms of the photon and
axion allow for a rotational $O(2)$ symmetry in the axion-photon field space.
  \item The interaction term, after dropping  a total time derivative, can also be expressed in 
an $O(2)$ symmetric way as follows:

\end{enumerate}

\begin{equation}
\label{axion photon int2}
S_I =  \frac{1}{2} \int dydzdt 
\beta \left[ \phi \pa_{t} A - A \pa_{t}\phi \right].
\end{equation}

The axion photon symmetry is (in the infinitesimal limit)

\begin{equation}
\label{axion photon symm}
\delta A = \epsilon \phi, \delta \phi = - \epsilon A,
\end{equation}

where $\epsilon$ is a small number. Using Noether`s theorem, this leads to the 
conserved current $j_{\mu}$, with components given by

\be
j_{0} = A \pa_{t}\phi - \phi \pa_{t} A - \frac{\beta}{2}(A^{2} + \phi^{2} )
\lab{axion photon density}
\ee

and 

\be
j_{i} = A \pa_{i}\phi - \phi \pa_{i} A.
\lab{axion photon current}
\ee

Here $i= y, z$ coordinates. We define now the complex field $\psi$ as
\be
\psi = \frac{1}{\sqrt{2}}(\phi + iA),
\lab{axion photon complex}
\ee
we see that 
in terms of this complex field, the axion photon density takes the form
\be
j_{0} = i( \psi^{*}\pa_{t}\psi - \psi \pa_{t} \psi^{*}) -  \beta \psi^{*}\psi.
\lab{axion photon density complex}
\ee

We observe that ,to first order in $\beta$, (\ref{axion photon int2}) represents the
interaction of the magnetic field with the "axion photon density" \rf{axion photon density}, \rf{axion photon density complex} 
and also that this interaction has the same form as that of scalar QED with an external "electric " field to first order. In fact the magnetic field (or more precisely $\beta /2$) appears to play the role of external electric potential that couples to the axion photon density \rf{axion photon density},\rf{axion photon density complex} which appears then to play the role of an electric charge density. From this analogy one can obtain without effort the scattering amplitudes, just using the known results from the scattering of charged scalar particles under the influence of an external static electric potential (see for example \cite{Bjorken-Drell}).

One should notice however that the natural initial states used in a real experiment, like an initial photon and no axion involved, is not going to have a well defined axion photon charge in the second
quantized theory (although its average value appears zero), so the 
S matrix has to be presented in a different basis than that of normal QED . This is similar to the difference between working with linear polarizations as opposed to circular polarizations in ordinary optics, except that here we talk about polarizations in the axion photon space.
In fact pure axion and pure photon initial states correspond to symmetric and antisymmetric linear combinations of particle and antiparticle 
in the analog QED language. The reason these linear combinations are not going to be mantained in the presence on $B$  in the analog 
QED language, is that the analog external electric potential breaks the symmetry between particle and antiparticle and therefore will not
mantain in time the symmetric or antisymmetric combinations.

From the point of view of the axion-photon conversion experiments,
the symmetry (\ref{axion photon symm}) and its finite form, which is just
a rotation in the axion-photon space, implies a corresponding symmetry
of the axion-photon conversion amplitudes, for the limit $\omega >>m$.

In terms of the complex field, the axion photon current takes the form
\be
j_{k} = i( \psi^{*}\pa_{k}\psi - \psi \pa_{k} \psi^{*}). 
\lab{axion photon current complex}
\ee

\section{The Particle Anti-Particle Representation of Axions and Photons and their Splitting in an External Magnetic Field}

Now let us introduce the charge conjugation \cite{part-antipart} , that is,

\be
\psi \rightarrow \psi^{*}. 
\label{charge conjugation}
\ee

We see then, that the free part of the action  is indeed invariant under (\ref{charge conjugation}).
The $A$ and $\phi$ fields when acting on the free vacuum give rise to a photon and an axion respectivelly,
but in terms of the particles and antiparticles defined in terms of  $\psi$, we see that a photon is an antisymmetric combination of particle and antiparticle and an axion a symmetric combination, since

\be
\phi =\frac{1}{\sqrt{2}}(\psi^{*} +\psi), ~A= \frac{1}{i\sqrt{2}}(\psi - \psi^{*}),
\lab{part, antipart}
\ee

so that the axion is even under charge conjugation, while the photon is odd.
These two eigenstates of charge conjugation will propagate without mixing as long as no external magnetic field is applied.
The interaction with the extenal magnetic field is not invariant under (\ref{charge conjugation}). In fact,
under (\ref{charge conjugation}) we can see that
\be
S_I \rightarrow - S_I.
\lab{non invariance}
\ee

Therefore  these symmetric and antisymmetric combinations, corresponding to axion and photon are not going to be maintained in the presence of $B$  in the analog QED language, since the "analog external electric potential" breaks the symmetry between particle and antiparticle and therefore will not mantain in time the symmetric or antisymmetric combinations.  In fact if the analog external electric potential is taken to be a
repulsive potential for particles, it will be an attractive potential for antiparticles, so the symmetry breaking is maximal.

Even at the classical level these two components suffer opposite forces, so under the influence of an inhomogeneous magnetic field both a photon or an axion will be decomposed through scattering into their particle and antiparticle components, each of which is scattered in a different direction, since the analog electric force is related to the gradient of the effective electric potential, i.e., the gradient of the magnetic field, times the $U(1)$ charge which is opposite for particles and antiparticles.

For this effect to have meaning, we have to work at least in a 2+1 formalism \cite{splt}, the 1+1 reduction \cite{duality}, \cite{part-antipart} which allows motion only in a single spacial direction is unable to produce such separation, since in order to separate particle and antiparticle components we need at least two dimensions
to obtain a final state with particles and antiparticles going in slightly different directions.

This is in a way similar to the Stern Gerlach experiment in atomic physics
\cite{Stern Gerlach}, where different spin orientations suffer a different force proportional to the gradient of the magnetic field in the direction of the spin. Here instead of spin we have that the photon is a combination of two states with different $U(1)$ charge and each of these components will suffer opposite force under the influence of the external inhomogeneous magnetic field. Notice also that since particle and antiparticles are distinguishable, there are no interference effect between the two processes.

Therefore an original beam of photons will be decomposed through scattering into two different elementary particle and antiparticle components
plus the photons that have not undergone scattering. These two beams are observable, since they have both photon components, so the observable
consequence of the axion photon coupling will be the splitting by a magnetic field of a photon beam. This effect being however an effect of first order in the axion photon coupling, unlike the ``light shining through a wall phenomena''.

\section{The Localized Soliton Solutions}
\label{solitionsol}

Now we turn to consider a time dependent axion and electromagnetic field with propagation only along the $z$ direction and where a time independent magnetic field pointing in the $x$-direction is present. This field may have only a $z$ dependence, to be determined later. We want however that this field will take into account the back reaction of the time dependent axion and electromagnetic fields in a time averaged way.

Repeating shortly the manipulations of the previous sections, where we choose again to work in the temporal gauge and consider only the $x$ polarization of the electromagnetic (time dependent) fields (i.e $E_{x} = - 
\partial_{t}A$, where $A$ is the $x$ component of the vector potential), we write the interaction term as (after dropping a total time derivative)

\begin{equation}
\label{inter}
	S_{I} = -\int d^{4}x\left[gB_{x}(z)\phi E_{x}\right] = \frac{1}{2}\int dzdtgB_{x}(z)\left[\phi\partial_{t}A - A\partial_{t}\phi\right],
\end{equation}

where $B_x = -\pa_{z} A_y$, since taking also that $A_z$ depends only on $z$ leave us only with this contribution. The 1+1 dimensional effective action is

\begin{equation}
\label{ }
	S_{3} = \int dzdt\left[\frac{1}{2}\partial_{\mu}A\partial^{\mu}A + \frac{1}{2}\partial_{\mu}\phi\partial^{\mu}\phi - \frac{1}{2}m^{2}\phi^{2} + gB_{x}(z)\phi\partial_{t}A - \frac{1}{2}(\partial_{z}A_{y})^{2}\right], 
\end{equation}

so now we can discuss the eq. of motion for this magnetic mean field \cite{soliton}

\be
\pa_{z}(\frac{ig}{2}( \psi^{*}\pa_{t}\psi - \psi \pa_{t} \psi^{*}) + B_x(z)) = 0.
\lab{B eq.}
\ee

The same result can be obtained from the original equations instead of the averaged Lagrangian obtained under the assumption that the mean field $B_x(z)$ is time independent and there doing a time averaging procedure, using for example
that under such time averaging $\phi \pa_{t} A$ equals $\frac{1}{2}(\phi \pa_{t} A -\pa_{t}\phi A)$ (here, again, $A$ denotes the $x$ component of the vector potential).
Equation \rf{B eq.} can be integrated, giving

\be
\label{mean field solution}
B_x(z) = - \frac{ig}{2}( \psi^{*}\pa_{t}\psi - \psi \pa_{t} \psi^{*}) + B_0,
\ee

where $B_0$ is an integration constant. The constant $B_0$ breaks spontaneously the charge conjugation symmetry of the theory (\ref{charge conjugation}), which is equivalent to changing the sign of $A$, since in such transformation the first term in the RHS of (\ref{mean field solution}) changes sign. This would be required in order to leave the interaction term (\ref{inter}) in the action invariant. However, the second term will not change (since it is a constant). Also, in problems where $B_x$ is taken as an external field \cite{part-antipart}, the interaction automatically breaks this ``charge conjugation'' symmetry.

We now consider $\psi$ to have the following time dependence,

\be
\psi = \rho(z)\exp(-i \omega t).
\lab{time dep.}
\ee

We want to see now what is the equation of motion for $\rho(z)$, which we take as a real field. We begin with the general eq. for $\psi$

\be
\pa_{\mu}\pa^{\mu} \psi + igB_x(z)\pa_{0}\psi = 0.
\lab{complex eq.}
\ee

Inserting \rf{time dep.} into (\ref{mean field solution}) and the result into \rf{complex eq.}, we obtain

\be
\frac{d^{2}\rho(z)}{dz^{2}} + \frac{d V_{eff}(\rho)}{d \rho} = 0,
\lab{analog mechanical}
\ee

where $V_{eff}(\rho)$ is given by

\be
V_{eff}(\rho) = \frac{1}{2}(\omega^{2} -  \omega g B_0)\rho^{2} + \frac{1}{4}g^{2}\omega^{2}\rho^{4}.
\lab{effective potential}
\ee

Some comments are required on the nature and signs of the different terms.
One should notice first of all that this effective potential is totally dynamically generated and vanishes when taking $ \omega = 0 $. 
Concerning signs, all terms proportional to $\omega^{2}$
are positive, in fact although the $(g\omega)^{2}$ term is quartic in $\rho$, it has to be regarded as originating not from  an ordinary potential of the scalar field in the original action, but rather from a term
proportional to a  $g^{2}(- i( \psi^{*}\pa_{t}\psi - \psi \pa_{t} \psi^{*}))^{2}$, quadratic in time derivatives, which could have been obtained if we had worked directly with the action rather than with the equations of motion, replacing (\ref{mean field solution}) back into the action i.e., integrating out the $B_x$ field (generically replacing solutions back in the action does not give correct results, here the procedure  gives correct results provided a lagrange multiplier enforcing magnetic flux conservation of (\ref{mean field solution}) is added, but this does not affect the terms dicussed here). Such type of quadratic terms in the time derivatives give a positive contribution both in the lagrangian and in the energy density, unlike a standard (not of kinetic origin) potential, where the contribution to the lagrangian is opposite to that of their contribution in the energy density.

The only term which may not be positive is the $-\omega g B_0$ contribution. This term breaks the charge conjugation symmetry (\ref{charge conjugation}) which for a field of the form \rf{time dep.} means $\omega \rightarrow -\omega$.

We can in any case choose the sign of $\omega$ such that the $-\omega g B_0$ contribution is negative and choose big enough  $B_0$
(or  $\omega $ small enough) so that this term makes the first term in the effective potential negative.

Now we are interested in obtaining solutions where 
$B_x(z) \rightarrow B_0$ as $z \rightarrow  \infty$ and also as $z \rightarrow  -\infty$, which requires $\rho \rightarrow 0$ as $z \rightarrow  \infty$ and also as $z \rightarrow  -\infty$. Since the vacuum with only a constant magnetic field is a stable one \ct{Ansoldi}.

The solution of the equations \rf{analog mechanical} and \rf{effective potential} with such boundary conditions is possible if
$\omega^{2}  - \omega g B_0 < 0$. After solving these analog of the "particle in a potential problem" with zero ``energy'', so that the boundary conditions are satisfied, we find that $\rho$ is given by (up to a sign),

\be
\rho = \frac{(\sqrt{2(\omega g B_0 - \omega^{2}) })/g\omega}{cosh(\sqrt{\omega g B_0 - \omega^{2} }(z-z_0))},
\lab{the solution}
\ee

where $z_0$ is an integration constant that defines the center of the soliton.

Inserting \rf{the solution} and \rf{time dep.} into the expression for $B_x$ (\ref{mean field solution}) we find the profile for
$B_x$ as a function of $z$. The difference in flux per unit length (that is ignoring the integration with respect to $y$ in the $yz$ plane) through the $yz$ plane of this solution with respect to the background solution
$B_x=B_0$ is finite amount. Since magnetic flux is conserved, we take this as an indication of the stability of this solution towards  decaying into the $B_x=B_0$ stable "ground state". 

Notice also that the soliton is charged under the $U(1)$ axion photon duality symmetry (\ref{axion photon symm}) and the vacuum is not, another evidence for the stability of these solitons. However for any given soliton,
there is no "antisoliton", since the condition $\omega^{2} -  \omega g B_0 < 0$ will not be mantained if we reverse the sign of $\omega$.
This is due to the fact that the vacuum of the theory, i.e. $B_x=B_0$ spontaneously breaks the charge conjugation symmetry (\ref{charge conjugation}).

There are some similarities, but also some crucial differences with the one dimensional topological solitons considered in Ref. \ct{T.D. Lee},  where also a complex field with a $U(1)$ global symmetry was considered and a time dependence of the form \rf{time dep.}
as well, which means that the soliton is charged under this $U(1)$ global symmetry, as in our case. The difference is in the type of potential
used and the origin of the potential. Here, the complete potential appears when considering non vanishing $\omega$, while in Ref. \ct{T.D. Lee}, 
 there is an original potential and self interaction in the original action and these self interactions appear with a negative 
sign in the effective potential, contrary to our case where the quartic self interaction enters with a positive sign (because of its kinetic origin as we explained before). Finally, in our case there is also a topological aspect as well, absent in Refs. \ct{T.D. Lee}, which is that the magnetic flux of the soliton differs from that of the vacuum.

\section{Summary and Conclusions}

When considering the scattering of axions and photons with the geometry relevant to the
axion-photon mixing experiments, the limit of zero axion mass reveals a continuous axion photon duality symmetry. This symmetry leads to a conserved current and then one observes that the interaction of the external magnetic field 
with the axion and photon is, to first order in the magnetic field, of the form of the first order in coupling constant 
interaction of charged scalars with an external electric scalar potential. Here the role of this ficticious external 
electric scalar potential is played (up to a constant) by the external magnetic field. 

Pure axion and pure photon initial states correspond to symmetric and antisymmetric linear combinations of particle and antiparticle 
in the analog QED language. Notice in this respect that charge conjugation of \rf{axion photon complex} corresponds to sign reversal of the photon field. The reason these linear combinations are not going to be maintained in the presence on a nontrivial $B$ in the analog 
QED language, is that the analog external electric potential breaks the symmetry between particle and antiparticle and therefore will not
mantain in time the symmetric or antisymmetric combinations.  

In this paper we presented the 2+1 dimensional generalization of a previous work that allowed only 1+1 reductions \ct{duality}, \ct{part-antipart} . One possible application of this that has not been discussed here could be the generalization of the soliton solutions found in Sec. \ref{solitionsol}.

We have focused now on the implications of representing a photon (and also the axion) as a linear combination of particle and antparticle.
Even at the classical level these two components suffer opposite forces, so both a photon or an axion under the influence of an inhomogeneous
magnetic field (since the analog electric force is related to the gradient of the effective electric potential, i.e., the gradient of the magnetic field) will be decomposed through scattering into its particle and antiparticle components, each of which is scattered in a different direction. For this effect to have meaning, we have to work at least in a 2+1 formalism, since the 1+1 reduction \ct{duality}, \ct{part-antipart}, which allows motion only in a single spatial direction, is unable to describe such separation, since in order to separate particle and antiparticle components we need at least two dimensions (in order to obtain a final state with particles and antiparticles in slightly different directions). Notice also that since particle and antiparticles are distinguishable, there are no interference effect between the two processes.

Therefore an original beam of photons will be decomposed through scattering into two different elementary particle and antiparticle components
plus the photons that have not undergone any scattering. These two beams are observable, since they have both photon components, so the observable
consequence of the axion photon coupling will be the splitting by a magnetic field of a photon beam. This effect being however an effect of first order in the axion photon coupling, unlike the  "light shining through a wall phenomena ". In this process one should account for "decoherence"
of the particle and antiparticle once they are well separated. As we see, beyond the question of observability, the existence of this kind of effects highlights many basic features of the axion photon system.

This resembles the Stern Gerlach experiment in atomic physics
\ct{Stern Gerlach},  where different spin orientations suffer a different force proportional to the gradient of the magnetic field in the direction of the spin. Here instead of spin we have that the photon is a combination of two states with different $U(1)$ charge and each of these components will suffer opposite force under the influence of the external inhomogeneous magnetic field.

Our analysis will apply also to neutral pions and photons, for example a very energetic gamma ray scattering from an inhomogeneous magnetic field
could give rise to two scattered beams (each of them containing both pions and photons) if the scattering takes place in the plane orthogonal to the magnetic field. Possible observable effects of photon splitting out of cosmic magnetic fields and not just laboratory ones could also be considered as a new source of multiple images for example.

We have also reviewed here the soliton solutions in the 1+1 dimensional case. In the context of a theory of containing a pseudo  scalar particle coupled to an electromagnetic field in the form
$g \phi \epsilon^{\mu \nu \alpha \beta}F_{\mu \nu} F_{\alpha \beta}$ an external constant magnetic field provides a stable vacuum \ct{Ansoldi}, unlike a constant electric field \ct{Ansoldi}. This can be understood qualitatively since a magnetic field is protected from decay to a lower energy density
configuration because the magnetic field cannot be screened, while this is not the case for a constant electric field. We  then studied a special type of geometry, where all the space dependence is on only one dimension, which we call $z$ and with fast and slow variables. The fast variables being the axion field and the vector potential in the $x$ direction,the slow variable, the magnetic field in the $x$ direction or the $y$ component of the vector potential, this is our ``mean field''. We then consider time averaging in the equations of motion or in the action so that the mean field is taken to be time independent and then considering the limit of zero axion mass we obtain a continuous axion photon ``duality symmetry'', and conserved quantities associated.

Introducing complex variables makes the structure of the time averaged equations quite manageable and the planar localized (in the $z$ direction) axion photon solutions are then found. From the profile for
$B_x$ as a function of $z$, we see that there is a difference in the flux per unit length (that is ignoring the integration with respect to $y$ in the $yz$ plane) through the $yz$ plane of this solution with respect to the background solution $B_x=B_0$. Since magnetic flux is conserved, we take this as an indication of the stability of this solution towards  decaying into the $B_x=B_0$ stable ``ground state''.

In future research it would be interesting to study generalizations of these solutions and look at the possibiliy of localizing not only with respect to one dimension (here $z$) but with respect to two. Another area of future research could be the consideration of a vacuum with charge density, which changes the discussion of solitons \ct{Bekenstein}. Finally one should also consider the effect of a small axion mass on these solutions.

It is important to mention that physically these solutions represent a new way of "trapping light", since photons participate in the solution in addition to axions. Also, from the form of the solution \rf{the solution}, we see that the critical magnetic field for the solution to be possible (at a fixed $\omega$) goes as $1/g$, but once the solution is possible, the strength of the soliton is also enhanced by a small $g$, so 
if these solitons are realized in the cosmos they could provided regions of high density trapped radiation that could lead to observable consequences due to interaction of this radiation with cosmic rays, or if the  soliton is suddenly destroyed (for example by a collision with another such type soliton) this could be the origin of a burst of radiation.

Another subject for future research should be the study of the effect of small axion mass.

Finally, an axion field or other psedoscalar interacting with non abelian fields through to the topological density and considering the presence of background chromomagnetic fields in the vacuum \ct{Nielsen} could may be introduce, through a suitable generalization of the soliton solutions found here, new features for the ground state of QCD. This could be the existence of gluon axion solitons.

\vskip.3in

\centerline{{\bf Acknowledgments}} 
We would like to thank Keith Baker, Doron Chelouche, Pierre Sikivie and Konstantin Zioutas for conversations and encouragement. 

\vskip.4in


\end{document}